\documentclass[prb,twocolumn,showpacs,amsmath,amssymb]{revtex4-1}
\usepackage{amsmath}
\usepackage{graphicx}
\usepackage{verbatim}
\usepackage{color}
\usepackage{subfigure}
\usepackage{hyperref}
\usepackage{soul}
\usepackage{amssymb}
\begin{document}
\title{Ginzburg-Landau theory of the zig-zag transition in quasi-one-dimensional classical Wigner
 crystals}
\author{J. E. \surname{Galv\'an-Moya}}
\author{F. M. \surname{Peeters}}
\email[Email: ]{francois.peeters@ua.ac.be}
\affiliation{Department of Physics, University of Antwerp, Groenenborgerlaan 171, B-2020,
 Antwerpen, Belgium}
%
\begin{abstract}
We present a mean-field description of the zig-zag phase transition of a quasi-one-dimensional
 system of strongly interacting particles, with interaction potential $r^{-n}e^{-r/\lambda}$,
 that are confined by a power-law potential ($y^{\alpha}$).  The parameters of the resulting
 one-dimensional Ginzburg-Landau theory are determined analytically for different values of
 $\alpha$ and $n$.  Close to the transition point for the zig-zag phase transition, the scaling
 behavior of the order parameter is determined.  For $\alpha=2$ the zig-zag transition from a
 single to a double chain is of second order, while for $\alpha>2$ the one chain configuration is
 always unstable and for $\alpha<2$ the one chain ordered state becomes
 unstable at a certain critical density resulting in jumps of single particles out of the chain.
\end{abstract}
\maketitle
%
\section{Introduction}

During the last two decades the interest in self-organized systems has increased enormously both
 experimentally and theoretically, due to its importance in solid-state
 physics, plasma physics as well as in atomic physics.  Wigner crystals are an elementary example
 of self-organization which has been realized in very diverse systems as e.g. electrons on liquid
 helium\cite{022_ikegami}, by using the well-known Paul and Penning
 traps\cite{020_itano,021_waki} to confine ions in a limited region, in dusty
 plasma\cite{033_dubin}, and more recently using static and radio-frequency electromagnetic
 potentials where crystallization was realized through laser cooling\cite{023_mortensen}.
 Additionally it has been proposed that these structures can be used for a possible
 implementation of a scalable quantum information processor\cite{024_cirac,026_leibfried}, and
 as quasi-one-dimensional (Q1D) Wigner crystals\cite{025_taylor}.  The theoretical analysis of
 these crystal structures has been realized previously, for 3D\cite{005_cornelissens,008_hasse},
 2D\cite{002_schweigert,004_schweigert,009_partoens} and Q1D\cite{003_piacente,014_fishman}
 systems.  From those studies it was shown that structural phase transitions can be induced by
 varying the strength of the external confinement potential and/or the density of particles.

In the present paper we concentrate on the ordered state of identical particles that are confined
 in a Q1D channel.  When the particles move in 2D and are confined by a
 parabolic\cite{003_piacente} or hard wall\cite{034_yang} potential in one of the in-plane
 directions the particles arrange themselves in parallel chains at low temperature.  Previously,
 it was found\cite{003_piacente} that with increasing density (or decreasing strength of the
 confinement potential) the system passes through a sequence of first and second order phase
 transitions where at each point the number of chains changes.  Of particular interest to us is
 the one chain to two chain transition which for a parabolic confinement potential was found to
 be a second order phase transition which occurs as a zig-zag transition.  Such a transition was
 observed experimentally\cite{030_valkering,031_chan,028_lutz,029_lutz} in systems with a finite
 number of particles, and the effects of a narrow channel and finite size of the system on the
 diffusion was recently analyzed in Refs. [\onlinecite{032_lucena}] and 
[\onlinecite{027_saintjean}].  Recently it was found theoretically\cite{006_piacente} that the
 analytic form of the confinement potential is very important for the occurrence of the zig-zag
 transition and the order of the phase transition.  Therefore, in the present paper we generalize
 the previous analysis to an arbitrary power law confinement potential (i.e. $y^{\alpha}$) and
 also to arbitrary inter-particle interaction which we model by $r^{-n}e^{-r/\lambda}$, which
 simulates most of the relevant experimental particle-particle interactions.  With this model
 potential we can simulate both short range and long range interactions.

We are interested in the behavior of the system at the zig-zag transition i.e. at the critical
 point.  This can be viewed as a spontaneous symmetry breaking and we will cast the problem into
 a mean field theory based on Landau's theory of phase transitions.  In this way we will
 construct a Ginzburg-Landau theory, for the single to two chains transition in a
 quasi-one-dimensional system of interacting particles.  We generalize the approach of Refs.
 [\onlinecite{014_fishman}] and [\onlinecite{011_delcampo}] to arbitrary power-law confinement
 and inter-particle interaction potential.  We obtain a Ginzburg-Landau equation for the order
 parameter close to the transition point, and determinate all the relevant parameters in this
 equation.  The order parameter is the distance of the particles from the trap axis.  By
 considering a large number of particles and using the local density approximation, we can
 consider the crystal as a continuum, so that the order parameter becomes a field.

The present paper is organized as follows. In Sect. \ref{theoretical_famework} we describe the
 model system and using Landau theory we find the behavior of the system close to the transition
 point.  Next we derive a Ginzburg-Landau equation for the system finding the dispersion
 relation.  In Sect. \ref{results} the results for the critical point and the normal mode
 spectrum are discussed.  Our conclusions and a discussion of possible quantum effects are given
 in Sect. \ref{conclusions}.

\section{Theoretical Framework}\label{theoretical_famework}

We consider a two-dimensional system consisting of $N$ particles with mass $m$ and charge $q$,
 which are allowed to move in the $x-y$ plane. The charged particles interact through a
 repulsive interaction potential; they are free to move in the $x$ direction but are confined by
 a one-dimensional potential which limits their motion in the $y$ direction. 
 The total energy of the system is given by:
\begin{equation}\label{total_energy}
    E = \frac{1}{2}m \sum_{i=1}^{N}{\mathbf{\dot{r}}_{i}^2} + V_{conf} + V_{int},
\end{equation}

\noindent where $V_{conf}$ and $V_{int}$ are the confinement and interaction potential respectively, given
 by
\begin{subequations}\label{Pot_Energy_original}
    \begin{equation}
        V_{conf} =
	    \frac{1}{2}m\upsilon_{t}^2 R^2\sum_{i=1}^{N}{\frac{|y_{i}|^{\alpha}}{R^\alpha}},
    \end{equation}
    \begin{equation}
        V_{int} = \sum_{i=1}^{N}\sum_{j>i}^{N} V_{pair}(r_{ij}),
    \end{equation}
\end{subequations}

\noindent where $V_{pair}(r_{ij})$ represents the inter-particle interaction.  The latter one will
 be taken as a screened power-law potential as follows:
\begin{equation}\label{screened_power_pot}
    V_{pair} =
        \frac{q^2}{\epsilon R} \frac{R^{n} e^{-r_{ij}/\lambda}}{r_{ij}^{n}},
\end{equation}

\noindent where $r_{ij}=|\mathbf r_{i}-\mathbf r_{j}|$ represents the relative position
 between the $i$-th and the $j$-th particle, the exponent $n$ is an integer and $\epsilon$
 is the dielectric constant of the medium the particles are moving in.  In the above, $R$ is an
 arbitrary length parameter which we introduced to guarantee the right units.  The energy can be
 written in dimensionless form
\begin{equation}\label{Total_Energy_dimensionless}
    E = \sum_{i=1}^{\infty}{\mathbf{\dot{r}}_{i}^2}
      + \upsilon^2\sum_{i=1}^{N}{|y_{i}|^{\alpha}}
      + \sum_{i=1}^{N}{\sum_{j > i}^{N}
        {\frac{e^{-\kappa r_{ij}}}{ r_{ij}^{n}}}},
\end{equation}
\noindent with dimensionless frequency $\upsilon$ given by $\upsilon=\upsilon_{t}/\omega_0$,
 while $\omega_0$ measures the strength of the confinement potential and $t_0=1/\omega_0$ is the
 unit of time.  The energy is expressed in units of
 $E_0=\left(m\omega_0^2/2\right)^{n/(n+\alpha)}
 \left(q^2/\epsilon\right)^{\alpha/(n+\alpha)}R^{(2n-\alpha)/(n+\alpha)}$ and all distances are
 expressed in units of
 $r_0=\left(2q^2/m\omega_0^2\epsilon\right)^{1/(n+\alpha)} R^{(n+\alpha-3)/(n+\alpha)}$.
 Additionally, the dimensionless parameter $\kappa=r_0/\lambda$ represents the screening
 parameter of the potential.  Limiting cases of this interaction potential are: Yukawa potential
 ($n=1$), power-law potential ($\kappa=0$), Coulomb potential ($\kappa=0, n=1$) and dipole
 interaction ($\kappa=0, n=3$).  We introduce a dimensionless linear density $\eta$ defined as
 the number of particles per unit of length along the unconfined direction.

In Ref. [\onlinecite{006_piacente}] it was demonstrated that for $\alpha>2$ the one-chain
 configuration is not stable for any values of $\eta$ and $\upsilon$.  Only for $\alpha=2$ the
 system exhibits a continuous transition from the one-chain to the two-chain configuration (i.e.
 zig-zag transition) at a transition point defined by a critical density ($\eta_c$) or a critical
 frequency ($\upsilon_c$).  For $\alpha<2$ the ground state configuration of the particles is,
 below $\eta_c$ or above $\upsilon_c$, arranged in a single chain.  Beyond this critical
 point particles are expelled one by one from this chain to positions parallel to the chain.

For the special case of $\alpha=2$ and before the transition point, the particles crystallize
 around the minimum point of the confinement potential $V$, at the positions
 $\mathbf{r}_{i}^{linear}=(i/\eta) \mathbf{e_x}$ with $i$ an integer.  The stability of the
 linear chain along the $x$ axis requires a relative transverse trap frequency exceeding a
 threshold value $\upsilon_c$ or a linear density smaller than $\eta_c$.  At this critical point,
 the configuration has a structural instability, such that for  $\upsilon<\upsilon_c$ or
 $\eta>\eta_c$ the particles are organized in a zig-zag structure, ordered in two chains with
 equilibrium positions
 $\mathbf{r}_{i}^{zigzag}=(i/\eta)\mathbf{e_x}+(-1)^{i}(c/\eta)\mathbf{e_y}$, where
 $c$ is a real and positive constant, $d=c/\eta$ represents the distance of each particle from
 the confinement potential minimum and $D=2d$ indicates the lateral separation between the two
 chains.

\subsection{Landau Theory for the zig-zag transition}
For the case $\alpha=2$ and near the zig-zag regime we follow the Landau theory approach of Ref.
 [\onlinecite{006_piacente}], and expand the total potential energy of the system as a
 function of the order parameter $c$ in a polynomial i.e. 
 $V(c) = V_{1ch}-Ac^2+Bc^4$, where $V_{1ch}$ represents the potential energy for the one-chain
 configuration.  Minimizing the potential energy we obtain the condition
\begin{equation}
    2\sum_{j=0}^{\infty} {\left[n\eta^n + \kappa\eta^{n-1}(2j+1)\right] \frac{e^{-\kappa
    \frac{2j+1}{\eta}}}{\left(2j+1\right)^{n+2}}}
    - \frac{\upsilon^{2}}{\eta^2} = 0.
\end{equation}

From this equation we obtain the value of $\eta_c(\upsilon)$ or $\upsilon_c(\eta)$ at which the
 single chain configuration becomes unstable.  Considering $\eta_c$ and expanding the potential
 energy around this critical value, we find that the order parameter close to the transition
 point is given by
\begin{equation}\label{c_landau}
    c = Y(\upsilon:n,\kappa)|\eta-\eta_c|^{\frac{1}{2}},
\end{equation}

\noindent with $Y(\upsilon;n,\kappa)=\sqrt{Y_{B}Y_{C}-Y_{A}Y_{D}}/Y_{B}$ where 
$        Y_{A} = n\eta_c^{n}S_{2} + \eta_c^{n-1}\kappa S_{1}
		- \upsilon^{2}/2\eta_c^2$,
$        Y_{B} = 2n(n+2)\eta_c^n S_{4} + 2(2n+1)\eta_c^{n-1}\kappa S_{3}
                + 2\eta_c^{n-2}\kappa^2 S_{2}$,
$        Y_{C} = n^2\eta_c^{n-1}S_{2} + (2n-1)\eta_c^{n-2}\kappa S_{1}
                + \eta_c^{n-3}\kappa^2S_{0} + \upsilon^2/\eta_c^3$,
$        Y_{D} = 2n^2(n+2)\eta_c^{n-1}S_{4} + 2[n(3n+1)-1]\eta_c^{n-2}\kappa S_{3}
                + 2(3n-1)\eta_c^{n-3}\kappa^2 S_{2} + 2\eta_c^{n-4}\kappa^3 S_{1}$, with

\begin{equation}
    S_{k} = \sum_{i=0}^{\infty} \frac{e^{-\kappa\frac{2i+1}{\eta_c}}}{(2i+1)^{n+k}}.
\end{equation}

\begin{figure}
\begin{center}
\includegraphics[scale=0.32]{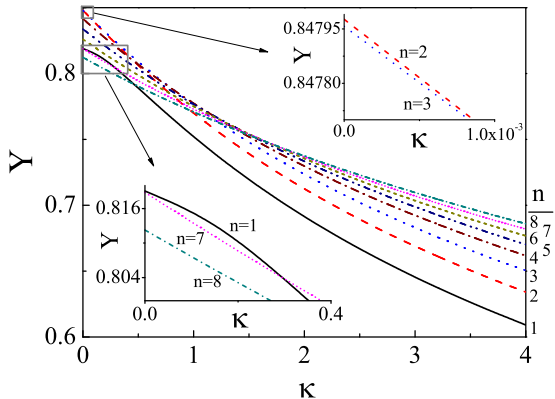}
\caption{\label{fig:y_kappa_landau} (Color online) The proportionality coefficient
 $Y(\upsilon;n,\kappa)$ as a function of $\kappa$ for different values of $n$ with $\upsilon=1$}.
\end{center}
\end{figure}

The value of $Y=Y(\upsilon;n,\kappa)$ is plotted in Fig. \ref{fig:y_kappa_landau} as function of
 $\kappa$ for different values of $n$ and a fixed value $\upsilon=1$.  Notice that the curves
 $Y(n,\kappa)$ for different values of $n$ cross each other at some value of $\kappa$.  The
 critical exponent of the order parameter of the zig-zag transition, Eq. (\ref{c_landau}), was
 verified experimentally on a low dimensional dusty plasma  in Ref. [\onlinecite{035_sheridan}]
 and theoretically in Refs. [\onlinecite{006_piacente}], [\onlinecite{014_fishman}],
 [\onlinecite{036_closson}] and [\onlinecite{037_sheridan}].

\subsection{Ginzburg-Landau Lagrangian for the zig-zag phase transition}\label{subsec_GLT}
Recently, this zig-zag transition (for $\alpha=2$, $\kappa=0$ and $n=1$) was cast into a
 mean-field description resulting in similar expressions as in the Landau theory of phase
 transitions.  This resulted in a one-dimensional Ginzburg-Landau type non-linear field
 theory\cite{011_delcampo}.  Here, we will extend the previous calculation
 to the more general problem described by the energy Eq. (\ref{Total_Energy_dimensionless}).  We
 start by considering the system in the situation that the one-chain configuration is stable but
 that it is close to the transition point.  The equilibrium positions of all the particles are
 along the $x$-axis.  We consider small oscillations around the equilibrium position of each
 particle as follows $x_{i}^{lin}=(i/\eta)+x_i$ and $y_{i}^{lin}=0+y_i$.  Then the relative
 position between the particles can be written as follows
\begin{equation}\label{position_Ate}
    r_{ij}= \left[ 1+ \frac{\tau_{ij}+\epsilon_{ij}}{A_{ij}}\right]^{1/2} (A_{ij})^{1/2},
\end{equation}

\noindent with $A_{ij}=(\Delta_{ij})^2$, $\tau_{ij}=2\Delta_{ij}(x_{i}-x_{j})$ and
 $\epsilon_{ij}=(x_{i}-x_{j})^2 + (y_{i}-y_{j})^2$, where ${\Delta_{ij}=(i-j)/\eta}$.
 Now, we assume that the vibration amplitudes in the axial and transverse direction are much
 smaller than the distance between the particles, i.e.
 $ \tau_{ij}, \> \epsilon_{ij}\ll A_{ij}$.  We expand Eq. (\ref{position_Ate}) and
 the exponential term as follows
\begin{eqnarray}
    r_{ij} & \approx & \Delta_{ij} + (x_{i}-x_{j}) \nonumber\\
     & &  + \left(1 
	  - \frac{(x_{i}-x_{j})}{\Delta_{ij}}
          + \frac{(x_{i}-x_{j})^2}{(\Delta_{ij})^2}
	  \right) \frac{(y_{i}-y_{j})^2}{2\Delta_{ij}},
\end{eqnarray}
\begin{eqnarray}\label{exp_Rk_approx}
    e^{-\kappa r_{ij}} \approx e^{-\kappa \Delta_{ij}}
        & &\left[ 1 - \kappa(x_{i}-x_{j}) + \kappa^2\frac{(x_{i}-x_{j})^{2}}{2} \right. \nonumber\\
        & &    - \left. \kappa\frac{(y_{i}-y_{j})^2}{2\Delta_{ij}}\right],
\end{eqnarray}

\noindent and similar the $n$-th power of the inverse of Eq. (\ref{position_Ate}) i.e. $1/r_{ij}^n$, as a
 Newton binomial around the equilibrium positions.  These expansions result in a decomposition of
 the total potential as
\begin{equation}\label{potential_after_expansion}
    V = \upsilon^2\sum_{i=1}^{\infty}{|y_{i}|^{\alpha}}
      + \sum_{l}^{\infty}V_{int}^{(l)},
\end{equation}

\noindent where the label $l$ indicates the order of the expansion.  Each order of the expansion of the
 interaction potential can be written as
\begin{equation}
    V_{int}^{(l)} = \frac{1}{2} \sum_{i\neq j}^{\infty} W_{ij}^{(l)},
\end{equation}

\noindent where the expansion terms up to fourth order are given by
\begin{widetext}
{\begin{eqnarray}
        W_{ij}^{(0)} & = & \frac{e^{-\kappa \Delta_{ij}}}{\Delta_{ij}^{n}}, \nonumber\\
        W_{ij}^{(1)} & = & -\frac{e^{-\kappa \Delta_{ij}}}{(\Delta_{ij})^{n+1}}
                            [\vartheta_{ij}n + \kappa\Delta_{ij}](x_{i}-x_{j}), \nonumber\\
        W_{ij}^{(2)} & = & \frac{e^{-\kappa \Delta_{ij}}}{2(\Delta_{ij})^{n+2}}
                            \left[
                            \left(n(n+1)-2\vartheta_{ij}n\kappa\Delta_{ij}
			    +\kappa^2\Delta_{ij}^2\right)(x_{i}-x_{j})^2
                            - \left(n+\kappa\Delta_{ij}\right)(y_{i}-y_{j})^2
                            \right], \nonumber\\
        W_{ij}^{(3)} & = & \frac{e^{-\kappa \Delta_{ij}}}{2(\Delta_{ij})^{n+3}}
                            n(x_{i}-x_{j}) \bigglb[ 
                            - \left(\frac{\vartheta_{ij}}{3}(n^2+3n+2)
                                    +(n+1)\kappa\Delta_{ij}
				    +\vartheta_{ij}\kappa^2\Delta_{ij}^2\right)(x_{i}-x_{j})^2
				    \nonumber\\ 
                     &   & +\left(\vartheta_{ij}(n+2) 
				    +\kappa\Delta_{ij}\right)(y_{i}-y_{j})^2
                            \biggrb],\nonumber\\
        W_{ij}^{(4)} & = & \frac{e^{-\kappa \Delta_{ij}}}{2(\Delta_{ij})^{n+4}} n\left[
                            \left(\frac{1}{12}(n^3+6n^2+11n+6)+3(n^2+3n+2)\kappa\Delta_{ij}
                                    + 2(n+1)\kappa^2\Delta_{ij}^2\right)(x_{i}-x_{j})^4
			    \right. \nonumber\\
                     &   & \left. -\left(\frac{1}{2}(n^2+5n+6)
				    +(9(n+2)\vartheta_{ij}+2(n+1))\kappa\Delta_{ij}
                                    +2\kappa^2\Delta_{ij}^2\right)(x_{i}-x_{j})^2(y_{i}-y_{j})^2
			   \right. \nonumber\\
                     &   & \left. +\left(\frac{1}{4}(n+2)+2\kappa\Delta_{ij}\right)
				     (y_{i}-y_{j})^{4}\right], \nonumber
\end{eqnarray}}
\end{widetext}

With ${\vartheta_{ij}=\Delta_{ij}/|\Delta_{ij}|}$.  It is sufficient to restrict
 ourselves to terms up to the fourth order and thus the potential can be written as 
$
 V_{int} \approx V_{int}^{(0)} + V_{int}^{(1)} + V_{int}^{(2)} + V_{int}^{(3)} + V_{int}^{(4)}
$.

\subsubsection{Representation in reciprocal space}\label{subsec_reciprocalspace}

Now we assume that the particles are pinned in the longitudinal direction and that they can only
 oscillate in the transverse direction ($x_i=0$).   Therefore, their normal axial modes can be
 neglected and we can discard the coupling to the longitudinal modes.  In this
 regime, we find that
{$        W_{ij}^{(0)} = e^{-\kappa \Delta_{ij}}/(\Delta_{ij}^{n})$},
{$        W_{ij}^{(1)} = 0$},
{$        W_{ij}^{(2)} = -\left(n+\kappa\Delta_{ij}\right)(y_{i}-y_{j})^2
			  e^{-\kappa\Delta_{ij}}/(2(\Delta_{ij})^{n+2}),
			  W_{ij}^{(3)} = 0$}, and 
{$        W_{ij}^{(4)} = n\left[(n+2)+8\kappa\Delta_{ij}\right](y_{i}-y_{j})^{4}
			  e^{-\kappa \Delta_{ij}}/(8(\Delta_{ij})^{n+4})$}.

In order to find the representation in reciprocal space, we define the normal
 modes of vibration in the transversal direction with wavevector $k$ as
 $\Psi_k=\psi^{(+)}_k-i\psi^{(-)}_k$ with amplitude $|\psi_k|^2=\psi^{(+)2}_k+\psi^{(-)2}_k$. 
 Following the standard process to find this representation as shown in e.g. Ref.
 [\onlinecite{014_fishman}] and using Plancherel's theorem\cite{B01_Yosida} for the confinement
 potential transformation, the different terms of the potential become
\begin{subequations}
    \begin{eqnarray}
        V_{conf} & = & \frac{\upsilon^2}{\sqrt{N}} \sum_{k>0} |\psi_k|^{\alpha}, \\
        V_{int}^{(1)} & = & 0, \\
        V_{int}^{(2)} & = & \sum_{k>0} \omega_{\perp}(n,\kappa,k)^2\psi_k^2, \\
        V_{int}^{(3)} & = & 0, \\
        V_{int}^{(4)} & = & \sum_{k_1+k_2+k_3+k_4=0} 
	      {A(k_1,k_2,k_3,k_4)\prod_{m=1}^{4}\psi_{_{k_m}}},
    \end{eqnarray}
\end{subequations}

\noindent where
 $	  \omega_{\perp}(n,\kappa,k)^2 = -\omega(n,\tilde\kappa,k)^2$,
 $	  A(k_1,k_2,k_3,k_4) = n\eta^{n+4}[(n+2)A_{0}(k_1,k_2,k_3,k_4)
				  + 8\tilde\kappa A_{1}(k_1,k_2,k_3,k_4)]/8$,
with $\tilde\kappa=\kappa/\eta$ and
\begin{subequations}
      \begin{eqnarray}\label{omega_k}
	  \omega(n,\tilde\kappa,k)^2 = 2\eta^{n+2} & \bigglb[ & 
		n\sum_{j=1}^{\infty} \frac{e^{-j\tilde\kappa}}{j^{n+2}}\sin^2\left(j\frac{k}{2}\right) \nonumber\\
	  & + & \kappa\sum_{j=1}^{\infty} \frac{e^{-j\tilde\kappa}}{j^{n+1}}\sin^2\left(j\frac{k}{2}\right)
	      \biggrb],
      \end{eqnarray}
      {
      \begin{eqnarray} \label{Acoef_Vpot4}
	  A_{0}(k_1,k_2,k_3,k_4) & = & \frac{2}{N}
		\sum_{j>0}{\frac{e^{-j\tilde\kappa}}{j^{n+4}}
		\prod_{m=1}^{4}{\sin\left(j\frac{k_m}{2}\right)}},\nonumber\\
	  A_{1}(k_1,k_2,k_3,k_4) & = & \frac{2}{N}
		\sum_{j>0}{\frac{e^{-j\tilde\kappa}}{j^{n+3}}
		\prod_{m=1}^{4}{\sin\left(j\frac{k_m}{2}\right)}},\nonumber\\
      \end{eqnarray}}
\end{subequations}

Due to the condition that the motion of the particles are restricted to the longitudinal
 direction it becomes apparent that the first and third order term of the interaction potential
 will be zero, and additionally we know that the first derivative equals zero because it is the
 necessary condition to have an equilibrium configuration.

\subsubsection{Minimum frequency of the interaction potential}\label{subsubsec_minfrequency}

From the definition of $\omega_{\perp}(n,\kappa,k)^2$ we find that its minimum value
 is located at $k_0=\pi$.  Lets expand for $k$ around this value ($k=k_0-\delta k$), and we
 obtain
\begin{subequations}
\begin{eqnarray}
    \omega_{\perp}(n,\kappa,k_0-\delta k)^2 & = & \omega_{\perp}(n,\tilde\kappa,k_0)^2 \nonumber\\
			  & & + h(n,\tilde\kappa)^2\delta k^2 
			  + {\footnotesize O(\delta k^4)}, \label{omega_perp_general} \\
    A(k_1,k_2,k_3,k_4) & = & \frac{1}{2N}\mathcal{A}(n,\tilde\kappa) + O(\delta k^2),
\end{eqnarray}
\end{subequations}

\noindent where $\omega_{\perp}(n,\tilde\kappa,k_0)^2 = -\varpi(n,\tilde\kappa)^2$ with
\begin{widetext}
\begin{subequations}
\begin{eqnarray}
    \varpi(n,\tilde\kappa)^2 & = &
        \left(\frac{\eta}{2}\right)^{n+2} e^{-\tilde\kappa}\left[
         2n\Phi\left(e^{-2\tilde\kappa}, n+2, \frac{1}{2} \right)
       + 4\tilde\kappa \Phi\left(e^{-2\tilde\kappa}, n+1, \frac{1}{2} \right)
      \right], \label{omega_perp_definition}\\
    h(n,\tilde\kappa)^2 & = & \left(\frac{\eta}{2}\right)^{n}
			\sum_{j=1}^{2}(-1)^{j+1} e^{-j\tilde\kappa} \left[
			     \frac{n}{2}\Phi\left(e^{-2\tilde\kappa},n,\frac{j}{2}\right)
			     + \tilde\kappa \Phi\left(e^{-2\tilde\kappa},n-1,\frac{j}{2}\right)
			\right], \label{h_alpha_0}\\
    \mathcal{A}(n,\tilde\kappa) & = & \left(\frac{\eta}{2}\right)^{n+4} n e^{-\tilde\kappa}\left[
				  \left(\frac{n}{2}+1\right)\Phi\left(e^{-2\tilde\kappa},n+4,\frac{1}{2}\right)
				+ 8\tilde\kappa\Phi\left(e^{-2\tilde\kappa},n+3,\frac{1}{2}
				  \right)
				\right],
\end{eqnarray}
\end{subequations}
\end{widetext}

\noindent where $\Phi(z,s,a)$ is the Lerch transcendent defined as
 $\Phi\left(z,s,a\right) = \sum_{k=0}^{\infty}  {z^k/\left(k+a\right)^s}$.
In Table \ref{tab:limit_GL_coeff} we show the limiting behavior of these terms.
It is important to note that the square of the transverse frequency is negative.
\begin{table*}
\caption{\label{tab:limit_GL_coeff} Behavior of the coefficients in the Ginzburg-Landau
 equation in two limiting cases, where $\tilde\kappa=\kappa/\eta$.}
\begin{tabular}{c|c|c}
\hline \hline
 & $\tilde\kappa \ll 1$ & $\tilde\kappa \gg 1$ \\
\hline
$\varpi(n,\tilde\kappa)^2$ &
 $2n \left(\frac{\eta}{2}\right)^{n+2} e^{-\tilde\kappa}
	      \sum_{j=0}^{\infty}\frac{1-2j\tilde\kappa}{(j+1/2)^{n+2}}$ &
 $2n\eta^{n+2} \tilde\kappa e^{-\tilde\kappa}$ \\
\hline
$ h(n,\tilde\kappa)^2$ &
 $\frac{n}{2} \left(\frac{\eta}{2}\right)^{n}
	      \sum_{j=0}^{\infty} \left(1-2j\tilde\kappa\right)
		\left(\frac{1}{(j+1/2)^{n}}-\frac{1}{(j+1)^{n}}\right)$ &
 $\left(\frac{\eta}{2}\right)^{n} \tilde\kappa e^{-\tilde\kappa}
	      \sum_{j=0}^{\infty} \frac{e^{-2j\tilde\kappa}}{(j+1/2)^{n-1}}$ \\
\hline
$\mathcal{A}(n,\tilde\kappa)$ &
 $n\left(\frac{n}{2}+1\right) \left(\frac{\eta}{2}\right)^{n+4}
	      \sum_{j=0}^{\infty}\frac{1-2j\tilde\kappa}{(j+1/2)^{n+4}}$ &
 $8n \left(\frac{\eta}{2}\right)^{n+4} \tilde\kappa e^{-\tilde\kappa}
	      \sum_{j=0}^{\infty}\frac{e^{-2j\tilde\kappa}}{(j+1/2)^{n+4}}$ \\
\hline \hline
\end{tabular}
\end{table*}

\subsubsection{Stability of the system}\label{subsubsec_cond_stability}

The system is stable when the second order term of the total potential energy (i.e. the
 coefficient of $\psi_k^2$) is minimum.  For the parabolic case ($\alpha=2$) the confinement
 potential term contributes to the second order of the total potential energy, then
\begin{equation}\label{Vtot_2}
    V^{(2)} = \sum_{k>0} \omega_{\perp}(n,\kappa,k)^2 \psi_k^2,
\end{equation}

\noindent where the transverse frequency is
\begin{equation}\label{beta}
    \omega_{\perp}(n,\kappa,k)^2 = \upsilon^2 - \omega(n,\tilde\kappa,k)^{2}
\end{equation}

\noindent and we note that its minimum value is reached for $k_0=\pi$.  Thus the critical value for the
 confinement frequency is given by
\begin{equation}\label{critical_frequency}
    \upsilon_c^2(n,\kappa) = \varpi(n,\tilde\kappa)^2.
\end{equation}

When $\upsilon>\upsilon_c$ the ground state configuration is a one-chain organization of the
 particles.  For $\upsilon<\upsilon_c$ the linear chain is unstable and the
 particles are arranged in a two-chain structure through a zig-zag organization.  When $\upsilon$
 is sufficiently close to the critical value $\upsilon_c$, an effective potential can be derived
 for the transverse normal modes $\psi_k$ with wavevector $\tilde k =k_0-\delta k$, such that
 $\delta k\ll 1$. The second order term of the effective potential is given by Eq.
 (\ref{Vtot_2}), where its coefficient, Eq. (\ref{omega_perp_general}), can now be written as
 follows
\begin{equation}\label{beta_closetomin}
    \omega_{\perp}(n,\kappa,k_0-\delta k)^2 = \delta_{\upsilon}(n,\tilde\kappa) 
			+ h(n,\tilde\kappa)^{2}\delta k^2,
\end{equation}

\noindent where $\delta_{\upsilon}(n,\tilde\kappa) = \upsilon^2-\upsilon_c(n,\tilde\kappa)^2$.  In the
 limiting case of a Coulomb inter-particle potential ($\kappa=0$, $n=1$) and considering
 $\eta=1$, we find
 $\upsilon_c(1,0) = \sqrt{7\zeta(3)}/2 = 1.45038$,
 $h(1,0) = \sqrt{log(2)/2} = 0.58871$ and
 $\mathcal{A}(1,0) = 93\zeta(5)/64 = 1.50679$,
\noindent which agrees with the results of Refs. [\onlinecite{011_delcampo}] and
 [\onlinecite{014_fishman}].  In Table \ref{tab:critical_values} we show the values of these
 terms for different interaction potentials.  Notice that the critical confinement frequency
 $\upsilon_c$ decreasing with increasing screening $\kappa$, and it increases with increasing
 density.  The relation between $\upsilon_c$ and $n$ depends on the density value, as will be
 discussed later.

\begin{table*}
\caption{\label{tab:critical_values} Values of the critical parameters for different values of
 $n$ and $\kappa$.}
\begin{ruledtabular}
\begin{tabular}{cc|ccc|ccc|ccc}
  &  &
 \multicolumn{3}{ c|}{$\eta=0.50$}&
 \multicolumn{3}{ c|}{$\eta=1.00$}&
 \multicolumn{3}{ c }{$\eta=1.50$} \\
\textrm{$n$}&
\textrm{$\kappa$}&
\multicolumn{1}{ c }{\textrm{$\upsilon_c(n,\tilde\kappa)$}}&
\multicolumn{1}{ c }{\textrm{$h(n,\tilde\kappa)$}}&
\multicolumn{1}{ c|}{\textrm{$\mathcal{A}(n,\tilde\kappa)$}}
&
\multicolumn{1}{ c }{\textrm{$\upsilon_c(n,\tilde\kappa)$}}&
\multicolumn{1}{ c }{\textrm{$h(n,\tilde\kappa)$}}&
\multicolumn{1}{ c|}{\textrm{$\mathcal{A}(n,\tilde\kappa)$}}
&
\multicolumn{1}{ c }{\textrm{$\upsilon_c(n,\tilde\kappa)$}}&
\multicolumn{1}{ c }{\textrm{$h(n,\tilde\kappa)$}}&
\multicolumn{1}{ c }{\textrm{$\mathcal{A}(n,\tilde\kappa)$}} \\
\colrule
 1 &  0.5 &    0.43113 &    0.38151 &    0.06332 &    1.36621 &    0.57569 &    2.13008 &    4.02285 &    0.82789 &    62.6067 \\
 1 &  1.0 &    0.31885 &    0.30221 &    0.04019 &    1.21941 &    0.53954 &    2.02615 &    3.86422 &    0.81415 &    68.1625 \\
 1 &  2.0 &    0.15131 &    0.15008 &    0.01002 &    0.90183 &    0.42740 &    1.28595 &    3.44901 &    0.76302 &    64.8368 \\
 2 &  0.5 &    0.37195 &    0.35188 &    0.06901 &    1.74736 &    0.80730 &    4.85728 &    7.52638 &    1.71598 &    299.492 \\
 2 &  1.0 &    0.26019 &    0.25394 &    0.04230 &    1.48781 &    0.70376 &    4.41649 &    6.98946 &    1.61460 &    310.865 \\
 2 &  2.0 &    0.11720 &    0.11676 &    0.01030 &    1.04076 &    0.50788 &    2.70688 &    5.95124 &    1.40752 &    282.655 \\
 3 &  0.5 &    0.30339 &    0.29542 &    0.05605 &    2.06259 &    0.99201 &    8.19081 &    12.7470 &    3.04303 &    1047.17 \\
 3 &  1.0 &    0.20566 &    0.20332 &    0.03331 &    1.71625 &    0.83557 &    7.17465 &    11.6677 &    2.80582 &    1048.42 \\
 3 &  2.0 &    0.08952 &    0.08936 &    0.00794 &    1.16342 &    0.57508 &    4.26315 &    9.70860 &    2.36334 &    918.354 \\
\end{tabular}
\end{ruledtabular}
\end{table*}

Additionally, from a simple expansion of the dispersion relation Eq. (\ref{beta}) around the
 equilibrium positions and for values of the frequency and density close to their critical
 values, the value of the parameter $c$ can be found from the non-linear algebraic equation
\begin{eqnarray}\label{c_harmonic}
    \upsilon^2 - 2\eta^{n+2} & \Bigglb[ & 
      n\sum_{j=1}^{\infty} 
	  \frac{e^{-\tilde\kappa\sqrt{(2j-1)^2+c^2}}}{[(2j-1)^2+c^2]^{\frac{n+2}{2}}} \nonumber\\
    & + & \tilde\kappa\sum_{j=1}^{\infty}
	  \frac{e^{-\tilde\kappa\sqrt{(2j-1)^2+c^2}}}{[(2j-1)^2+c^2]^{\frac{n+1}{2}}}
	   \Biggrb] = 0 
\end{eqnarray}

\subsubsection{Continuum approximation}\label{subsubsec_continuumapprox}

Close to the transition point ($\delta k\ll 1$), the transverse deviation of the particles is
 very small and we can use a continuum approach for these modes.  In doing so we replace the
 discrete sum over $\tilde k$ by an integral
 $\sum_{k} \rightarrow \int{d(\delta k)N/2\pi}$. 
Using the Fourier transform we obtain a continuous form for the modes $\psi(x)$ as follows
$\psi_k = \int{dx \psi(x) e^{-i\delta k x}/\sqrt{N}}$.
Then the remaining terms of the potential becomes
\begin{subequations}
      \begin{eqnarray}
	  V^{(\alpha)} = \frac{1}{2} & \displaystyle\int & \upsilon^2 \left|\psi(x)\right|^{\alpha} dx, \\
	  V^{(2)} = \frac{1}{2} & \displaystyle\int & \bigglb[-\varpi(n,\tilde\kappa)^2\psi(x)^2 \nonumber\\
	  & &  + {h(n,\tilde\kappa)}^2\left(\frac{\partial \psi}{\partial x}\right)^2 
	  \biggrb]dx, \\
	  V^{(4)} = \frac{1}{2} & \displaystyle\int & \mathcal{A}(n,\tilde\kappa)\psi(x)^4 dx.
      \end{eqnarray}
\end{subequations}

Finally, we obtain the Lagrangian $L=\int{\mathcal{L}(x)dx}$ where the Lagrangian density
 $\mathcal{L}(x)$ reads
\begin{widetext}
\begin{equation}\label{density_lagrange}
    \mathcal{L}(x)=
	  \frac{1}{2}\left[(\partial_t \psi(x))^2-h(n,\tilde\kappa)^2(\partial_x \psi(x))^2
          + \varpi(n,\tilde\kappa)^2\psi(x)^2
	  - \upsilon^2|\psi(x)|^{\alpha}
	  - \mathcal{A}(n,\tilde\kappa)\psi(x)^4 \right].
\end{equation}
\end{widetext}

\begin{figure*}
\begin{center}
\subfigure{
\includegraphics[scale=0.32]{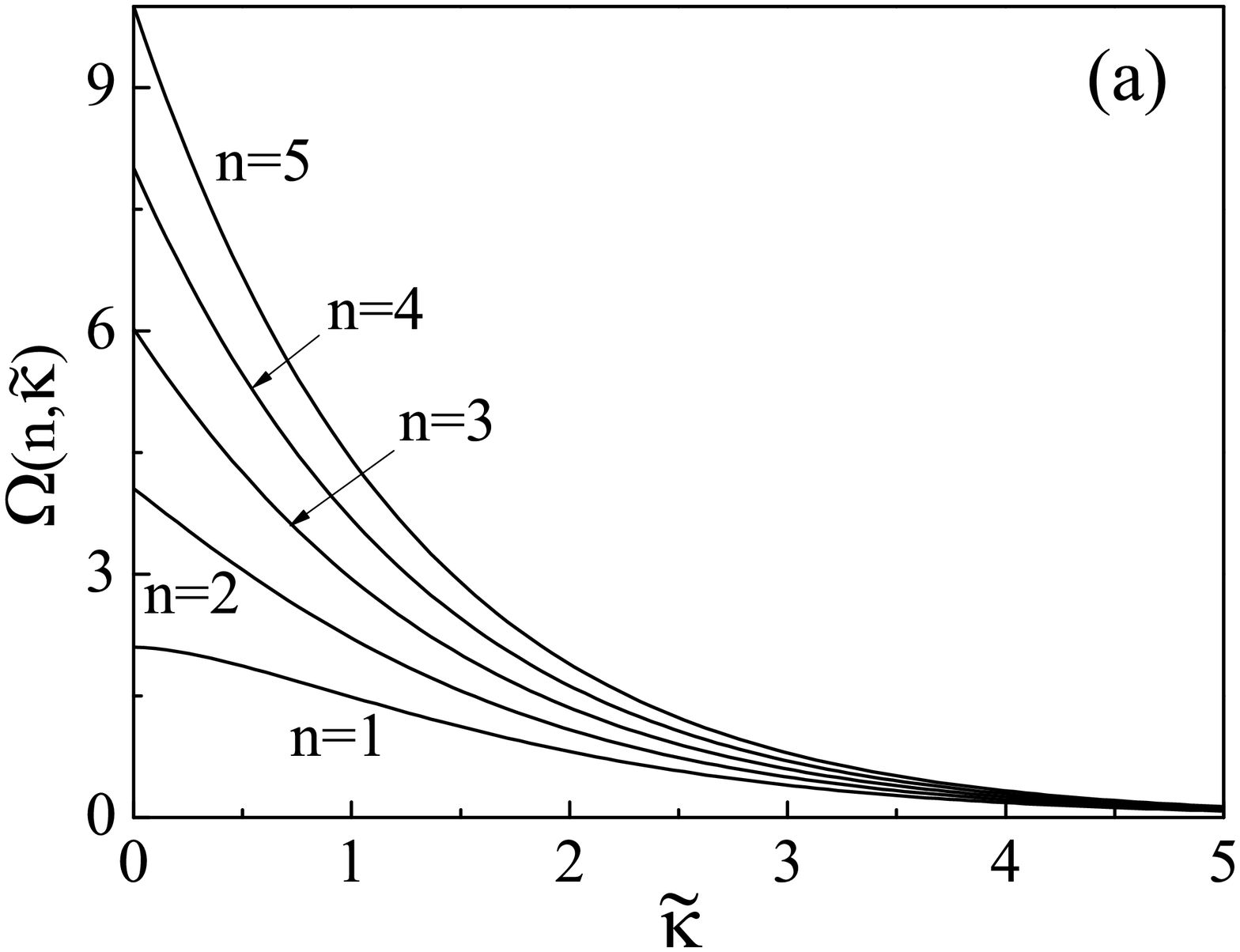}}
\subfigure{
\includegraphics[scale=0.32]{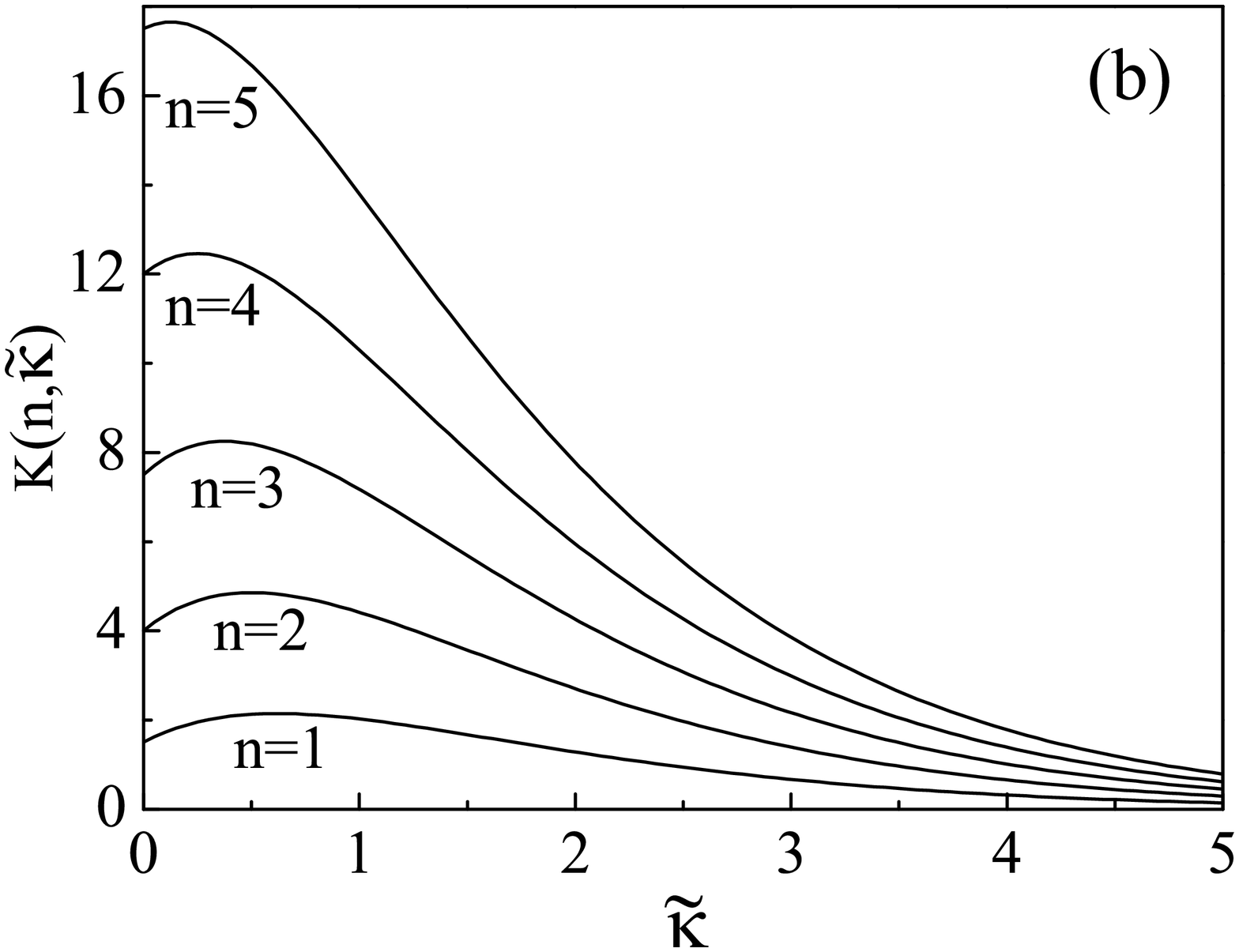}}
\caption{\label{fig:gl_coefficients} Ginzburg-Landau coefficients for the potential energy
 density (Eq. (\ref{pot_energy_density})) as a function of the screening parameter
 $\tilde\kappa=\kappa/\eta$, for different values of $n$.}
\end{center}
\end{figure*}

In the special case $\alpha=2$, $\kappa=0$, $n=1$ this Lagrangian density is the one found in
 Ref. [\onlinecite{011_delcampo}],
 and it has the form of a Ginzburg-Landau equation  (Refs.
 [\onlinecite{011_delcampo}] and [\onlinecite{B02_Landau}]).  Defining $\varphi(x)=\eta\psi(x)$
 and $\tilde\upsilon(n,\tilde\kappa)^2=\upsilon(n,\tilde\kappa)^2/\eta^{n+\alpha}$, we may find
 from Eq. (\ref{density_lagrange}) an expression for the potential energy density $\mathcal{V}$:
\begin{eqnarray}\label{pot_energy_density}
  \frac{2\mathcal{V}}{\eta^n} = & & K(n,\tilde\kappa)\varphi(x)^4
		+ \tilde\upsilon(n,\tilde\kappa)^2|\varphi(x)|^{\alpha} \nonumber\\
		& - & \Omega(n,\tilde\kappa)\varphi(x)^2,
\end{eqnarray}

\noindent with the real positive coefficients
 $\Omega(n,\tilde\kappa)=\varpi(n,\tilde\kappa)^2/\eta^{n+2}$ and
 $K(n,\tilde\kappa)=\mathcal{A}(n,\tilde\kappa)/\eta^{n+4}$ that are plotted in Fig.
 \ref{fig:gl_coefficients} as function of $\tilde\kappa$ for different values of $n$.  Notice
 that both coefficients are positive and decrease with increasing $\tilde\kappa$.  Now, the
 density $\eta$ plays the role of a scaling parameter in the potential energy density
 ($\mathcal{V}$), in the screening parameter ($\kappa$), in the strength of the confinement
 ($\upsilon$) and in the order parameter ($\psi$).  For $\alpha=2$ we find the usual Landau
 energy expression for a second-order phase transition
\begin{equation}
  \frac{2\mathcal{V}}{\eta^n} = K(n,\tilde\kappa)\varphi(x)^4
		+ [\tilde\upsilon(n,\tilde\kappa)^2 - \Omega(n,\tilde\kappa)]\varphi(x)^2.
\end{equation}

\subsubsection{Equation of motion}\label{subsubsec_eqofmotion}

From Eq. (\ref{density_lagrange}) we obtain the equation of motion for $\psi(x)$ as
 follows
\begin{widetext}
\begin{equation}\label{Ec_euler_lagrange_general}
    \partial_{t}^2\psi(x)
    - h(n,\tilde\kappa)^2 \partial_{x}^2\psi(x)
    - \varpi(n,\tilde\kappa)^2\psi(x)
    + \frac{\alpha\upsilon^2}{2} sign(\psi(x)) |\psi(x)|^{\alpha-1}
    + 2 \mathcal{A}(n,\tilde\kappa)\psi(x)^3 = 0.
\end{equation}
\end{widetext}

In this context the order parameter $\psi(x)$ represents a continuous version for the value of
 $c$ which is the distance of the particles ($d=c/\eta$) from the minimum of the
 confinement potential.  When the order parameter varies slowly in space, the time
 independent version of Eq. (\ref{Ec_euler_lagrange_general}) becomes
\begin{equation}\label{Ec_euler_lagrange}
    \varpi(n,\tilde\kappa)^2\psi
    - \frac{\alpha\upsilon^2}{2} sign(\psi) |\psi|^{\alpha-1}
    - 2 \mathcal{A}(n,\tilde\kappa)\psi^3 = 0.
\end{equation}

We note that for $\alpha<1$ a one-chain configuration is not allowed, because $\psi=0$ is not
 a solution of Eq. (\ref{Ec_euler_lagrange}) in this case.

Considering $\alpha \geq 1$ and defining
 $\tilde a(n,\tilde\kappa) = \alpha\upsilon^2/2\eta^{n+2}\Omega(n,\tilde\kappa)$ and
 $\tilde b(n,\tilde\kappa)=2\eta^2 K(n,\tilde\kappa)/\Omega(n,\tilde\kappa)$, the latter equation
 is reduced to
\begin{equation}\label{Ec_euler_lagrange_simplify}
    1 - \tilde a(n,\tilde\kappa)|\psi|^{\alpha-2} - \tilde b(n,\tilde\kappa) \psi^2 =0.
\end{equation}

For $\alpha=2$ this equation results in a second order transition, from the single
 chain (i.e. $\psi=0$) to the zig-zag (i.e. $\psi\neq 0$) configuration, with the critical
 point defined by $\tilde a(n,\tilde \kappa)=1$ which in fact is a generalization of Eq.
 (\ref{c_harmonic}).

\begin{figure}
\begin{center}
\includegraphics[scale=0.32]{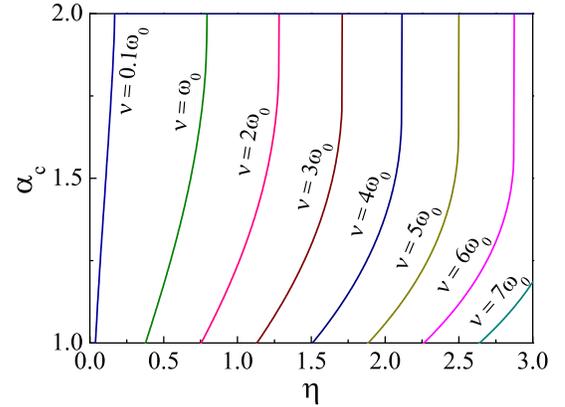}
\caption{\label{fig:critical_alpha} (Color online) Critical value of the exponent of the
 confinement frequency as a function of the density for different values of the strength of the
 confinement potential.  We took the parameters $n=1$ and $\kappa=1$}
\end{center}
\end{figure}

For $1<\alpha<2$ and minimizing Eq. (\ref{Ec_euler_lagrange_simplify}) we find 
\begin{equation} \label{Ec_trasc_critical_alpha}
    2\Omega(n,\tilde\kappa) -
      \left( \frac{\alpha_c^2 \upsilon^2}{2 \eta^{n+\alpha_c}} \right)^{\frac{2}{4-\alpha_c}}
      \left(\frac{4\alpha_c}{2-\alpha_c} K(n,\tilde\kappa)\right)^{\frac{2-\alpha_c}{4-\alpha_c}}
      = 0
\end{equation}

\noindent which represents a non-linear equation for the critical
 exponent of the confinement potential ($\alpha_c$), which is the minimum value of $\alpha$
 for which a one-chain configuration is the ground state configuration.  From Eq.
 (\ref{Ec_trasc_critical_alpha}) we note that this critical value will be at most equal to 2, as
 shown in Fig. \ref{fig:critical_alpha} for different values of the strength of the confinement
 frequency.

Finally we may find analytical expressions for the order parameter from Eq.
 (\ref{Ec_euler_lagrange_simplify}) for different values of $\alpha \geq 2$, which are given in
 Table \ref{tab:psi_zigzag}.  Notice that it is always possible to find a $\psi \neq 0$ which
 indicates that the single chain configuration is always unstable when $\alpha>2$.  For
 $\alpha=2$ we find $\psi \neq 0$ only when $\tilde a(n,\tilde\kappa)<1$.

\begin{table}
\caption{\label{tab:psi_zigzag} Order parameter for different values of
 $\alpha$.}
\begin{tabular}{c|c}
\hline \hline
 $\alpha$ & $\psi$ \\
\colrule
 2 & $\sqrt{(1-\tilde a)/\tilde b}$ \\
 3 & $\left(-\tilde a + \sqrt{\tilde a^2 + 4 \tilde b} \right)/2\tilde b $ \\
 4 & $1/\sqrt{\tilde a + \tilde b}$\\
\hline \hline
\end{tabular}
\end{table}

\section{Results and discussion}\label{results}

As has been found in previous section, a continuous zig-zag transition occurs for
 parabolic confinement.  For the case ($\alpha \neq 2$) it is not possible to define a transition
 between the one-chain and two-chain configuration, because the confinement potential does not
 contribute to the second order term of the total potential energy.  Additionally we find
 that the minimum value of the transverse frequency is purely imaginary, see
 Eqs. (\ref{omega_perp_general}) and (\ref{omega_perp_definition}), and this condition implies
 that in this case the transition for the one-chain to the two-chains configuration is not
 allowed which agrees with previous\cite{ 006_piacente} results.  We also performed Monte-Carlo
 simulations and found that for $\alpha > 2$ the one-chain configuration is never formed for any
 value of the density and the confinement frequency. However from similar simulations one can
 show that for $\alpha<2$ the one-chain configuration is stable until a critical point, beyond
 which the configuration is changed to a single chain containing vacancies due to
 jumps of individual particles away from the chain axis.

\subsection{Transition point for $\alpha \leq 2$}
\begin{figure}
\begin{center}
\includegraphics[scale=0.32]{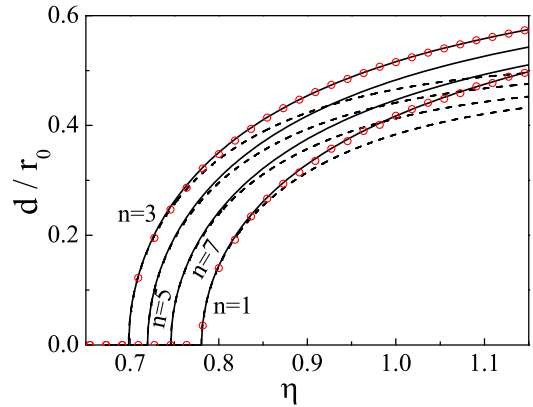}
\caption{\label{fig:c_density} (Color online) Displacement from the $x$-axis as a function of the
 linear density for $\upsilon=1$.  The dashed lines are obtained from the Landau theory while the
 solid lines are the solutions of Eq. (\ref{c_harmonic}). The open circles represent the results
 of our Monte-Carlo simulations for $n=1$ and $n=3$}. 
\end{center}
\end{figure}

\begin{figure}
\begin{center}
\includegraphics[scale=0.32]{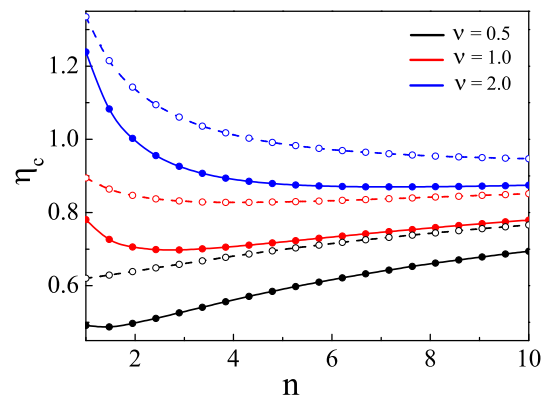}
\caption{\label{fig:etac_n} (Color online) Critical linear density as a function of the $n$
 exponent of the inter-particle interaction for different values of the parabolic confinement
 frequency.  The (solid and dashed) lines are the prediction of the Landau theory and the (solid
 and open) circles are found with the present method (for $\kappa=0$ and $\kappa=1$
 respectively).}
\end{center}
\end{figure}

For the case of a power-law inter-particle potential ($\kappa=0$) with parabolic confinement
 and using dimensionless units, it is possible to find an analytical relationship between the
 confinement frequency and the linear density as $\eta = \upsilon^{-2/(n+2)}$. For this
 case we show in Fig. \ref{fig:c_density} the behavior of the order parameter as a function of
 the linear density for different values of $n$.  Dashed curves represent the solution from the
 Landau theory, Eq. (\ref{c_landau}), the full curves are the solution of Eq. (\ref{c_harmonic})
 and they are compared with the results of a Monte-Carlo simulation for $n=1$ and $n=3$ (open
 circles in Fig. \ref{fig:c_density}).  From these results we notice that there is perfect
 agreement between our calculation and the exact results obtained from Monte-Carlo simulations.
 In the same context Fig. \ref{fig:etac_n} shows the variation of the critical density as a
 function of the exponent $n$ as obtained from Landau theory (solid and dashed curves for
 $\kappa=0$ and $\kappa=1$, respectively) and the result from the present work (full and open
 circles for $\kappa=0$ and $\kappa=1$) the results are shown for different confinement
 frequencies.  Notice that for $\upsilon=1$ this function has a local minimum where the dipole
 potential exhibits the lowest critical density.

\begin{figure}
\begin{center}
\includegraphics[scale=0.32]{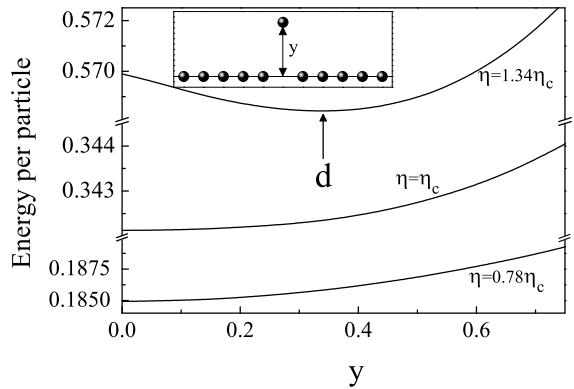} 
\caption{\label{fig:yfix_derivative} Potential energy per particle as a function of the position
 of one of the particles in the confinement direction, before ($\eta=0.78\eta_c$), after
 ($\eta=1.34\eta_c$) and at the transition ($\eta=\eta_c$) for parabolic confinement.  We took
 the parameters $\upsilon=1$, $\kappa=1$ and $n=1$}
\end{center}
\end{figure}

On the other hand, it is also possible to find the value of $d$ numerically by fixing one
 particle at a distance $y$ from the one-chain axis in the confinement direction and minimise the
 energy with respect to the position of the other particles.  The resulting minimum potential
 energy of the system is shown in Fig. \ref{fig:yfix_derivative} for a Yukawa inter-particle
 potential with $\kappa=1$ and $\upsilon=1$.  For $\eta\leq\eta_c$ the minimum is found at $y=0$
 and for $\eta>\eta_c$ it continuously shifts to $y\neq 0$, which is typical for a second order
 transition.

\begin{figure*}
\begin{center}
\subfigure{
\includegraphics[scale=0.3]{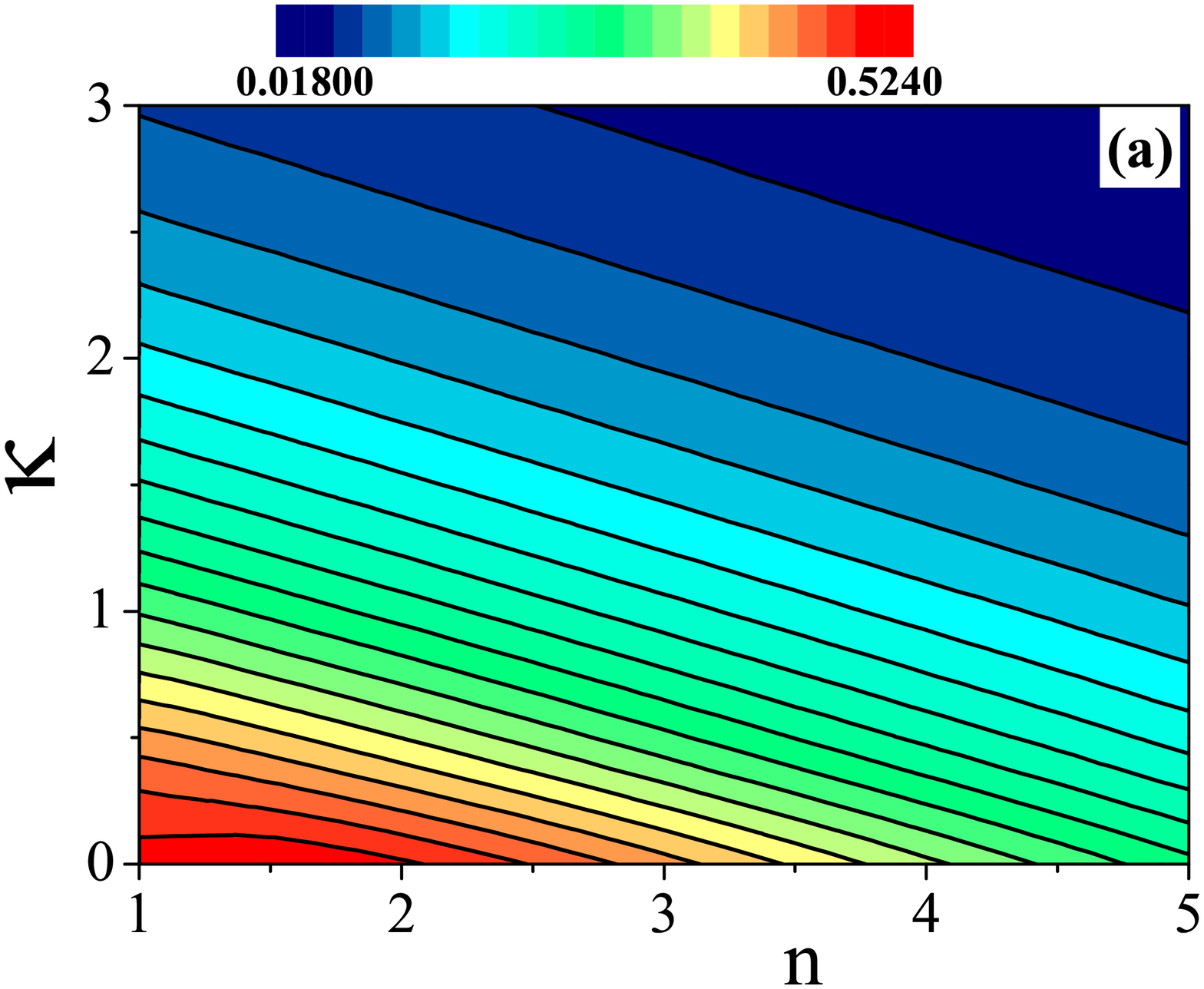}}
\subfigure{
\includegraphics[scale=0.3]{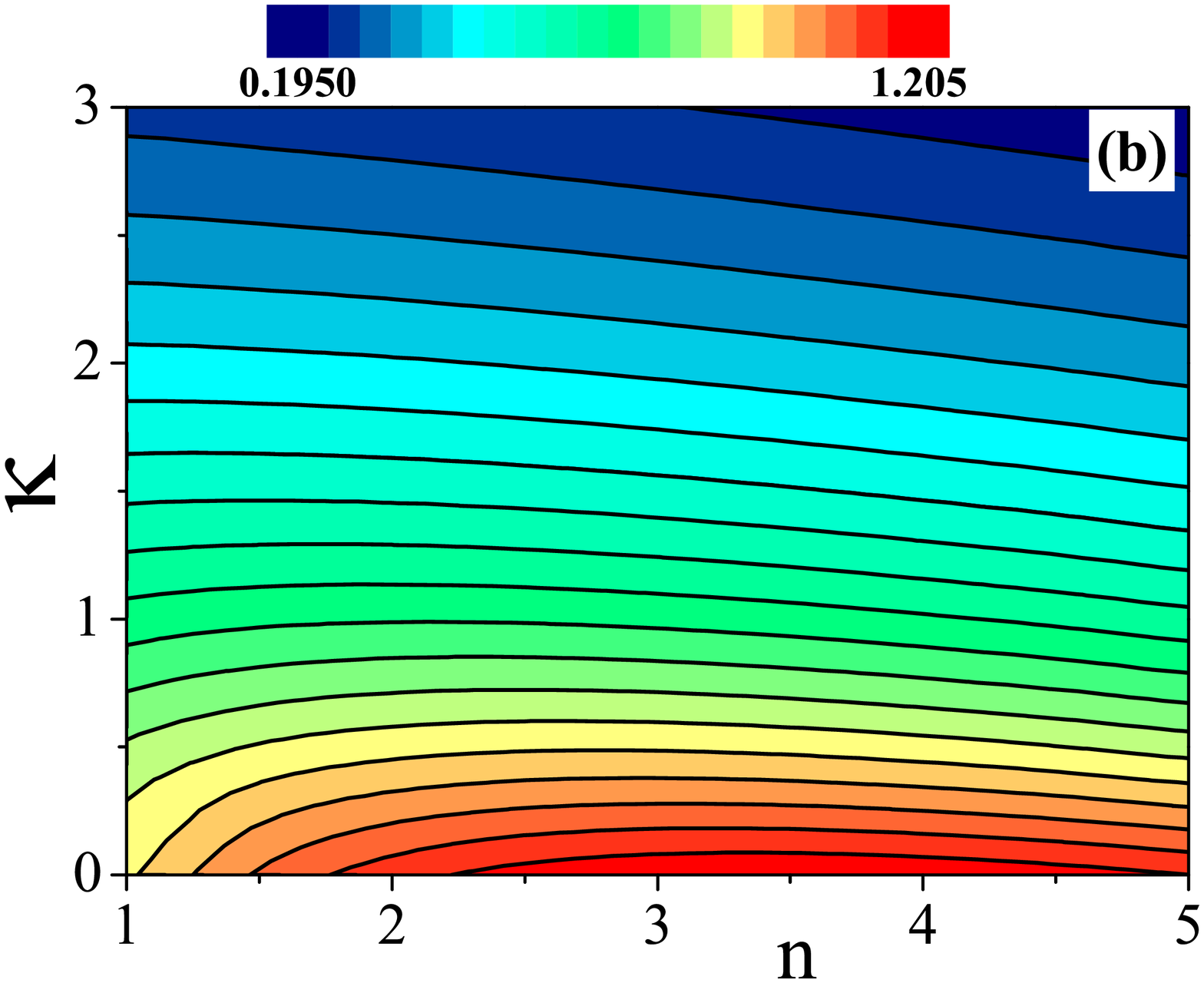}}
\subfigure{
\includegraphics[scale=0.3]{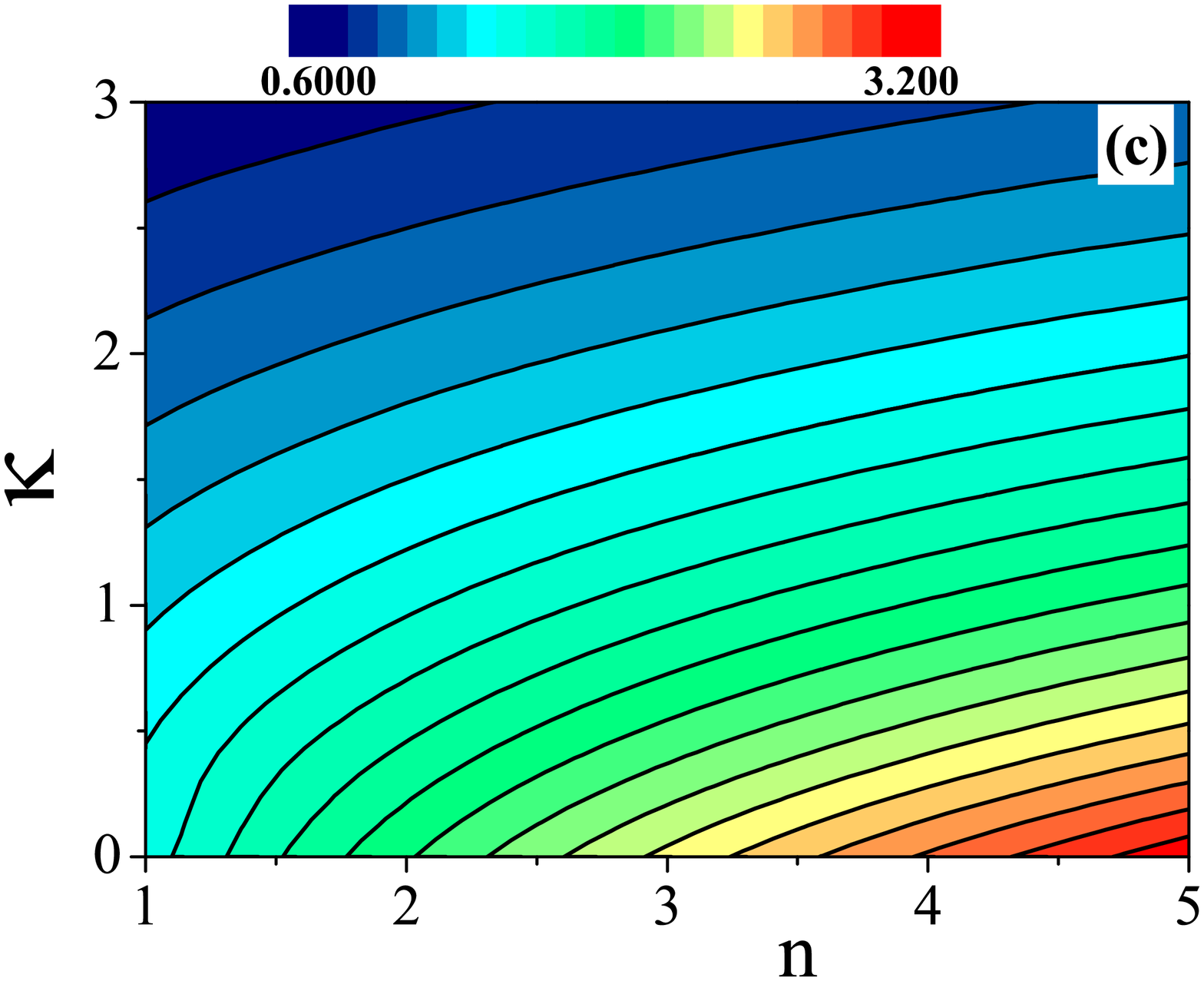}}
\subfigure{
\includegraphics[scale=0.3]{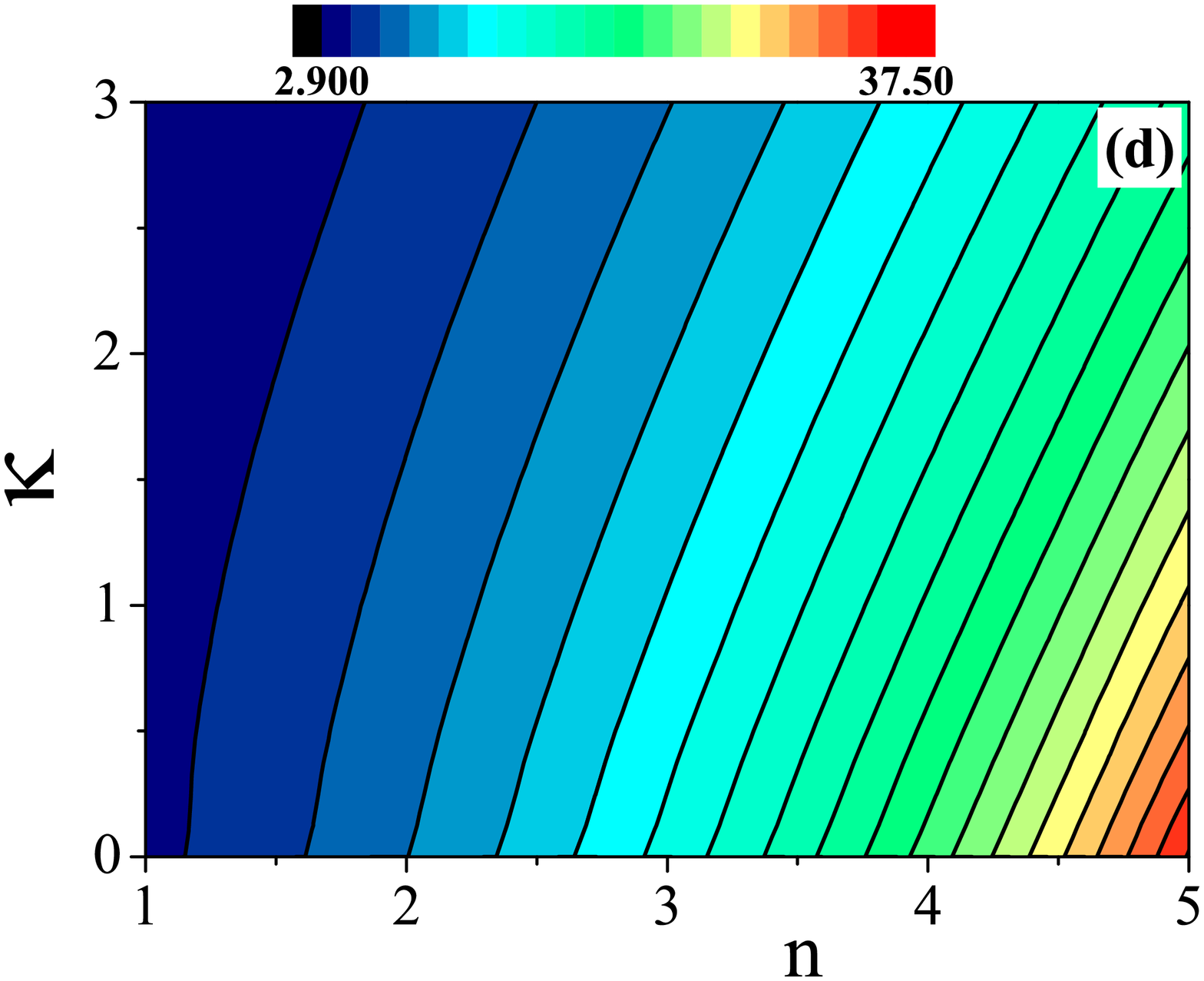}}
\caption{\label{fig:critical_density_contours} (Color online) Contour plots of the critical
 frequency $\upsilon_c$ as a function of $n$ and $\kappa$, for (a) $\eta=0.50$, (b) $\eta=0.75$, (c)
 $\eta=1.00$, (d) $\eta=2.00$, in case of a parabolic confinement potential.}first
\end{center}
\end{figure*}

From Eq. (\ref{critical_frequency}) we draw the contour plot of $\upsilon_c$ as a function of
 $n$ and $\kappa$ for several values of the density (when making the contour plot we replaced
 $n$ by a real number), which are shown in Fig. \ref{fig:critical_density_contours}.  We
 observe a strong dependence of the highest value of $\upsilon_c$ on $\eta$, and therefore the
 region of frequencies over which the one-chain configuration exists.  For low densities
 ($\eta<1$) the one-chain organization is dominant for small values of the exponent $n$
 and gradually this region is extended to higher values of $n$ with increasing $\eta$.  This
 result shows that for low densities the one-dimensional behavior of the system is a better
 representation for the Coulomb and dipole inter-particle potential.  For $\eta\geq 1$ the
 one-dimensional region of frequencies increasing with increasing $n$.  In all cases the
 critical frequency decreases with increasing $\kappa$.

\begin{figure*}
\begin{center}
\includegraphics[width=\textwidth]{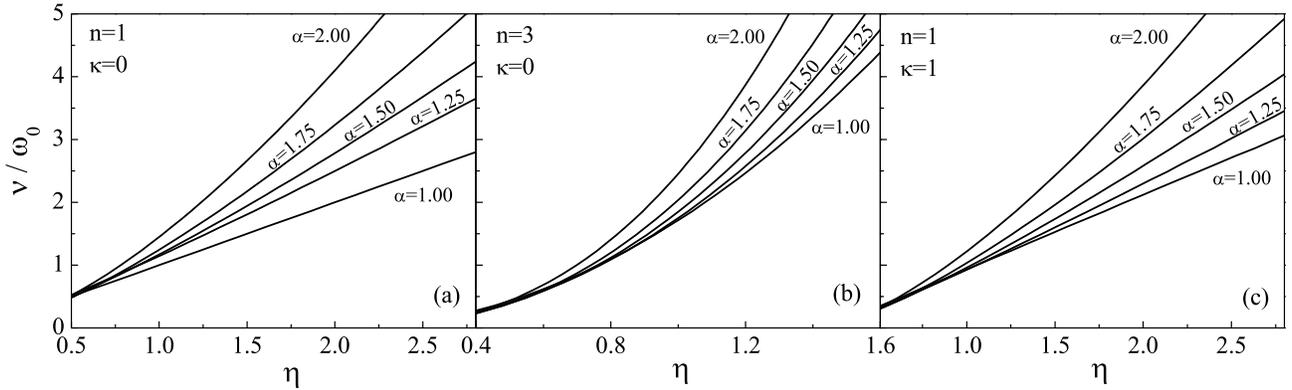}
\caption{\label{fig:critical_point} Critical transition point for the one-chain configuration.
 Above the curves the one chain configuration is the ground state.  Results are shown for (a)
 Coulomb potential, (b) dipole potential and (c) Yukawa potential, for different values of the
 exponent of the confinement potential $\alpha$.}
\end{center}
\end{figure*}

Our previous mean-field theory was derived for modes of the linear chain close to the instability
 point.  Therefore, it is possible to find the critical point for the aforementioned instability
 from Eq. (\ref{Ec_euler_lagrange}).  In Fig. \ref{fig:critical_point} we plot the
 transition point at which the one-chain structure becomes unstable for $\alpha \leq 2$.  Above
 each curve only $\psi=0$ is a solution of Eq. (\ref{Ec_euler_lagrange}).  Only for
 $\alpha=2$ the curves corresponds to a second order zig-zag transition.  Notice that the
 stability region for the single chain configuration increases with decreasing $\alpha$.

\begin{figure*}
\begin{center}
\subfigure{
\includegraphics[scale=0.32]{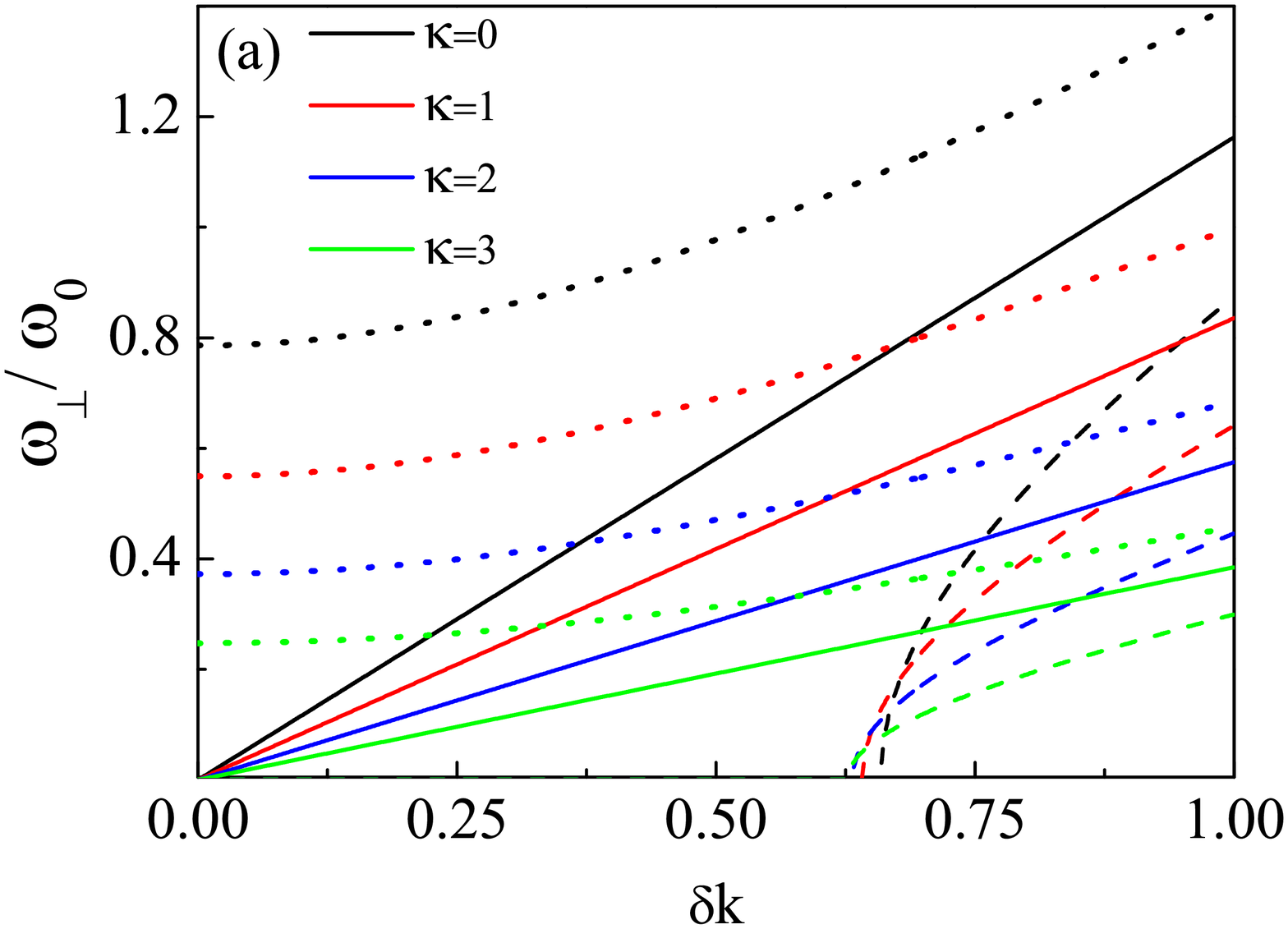}}
\subfigure{
\includegraphics[scale=0.32]{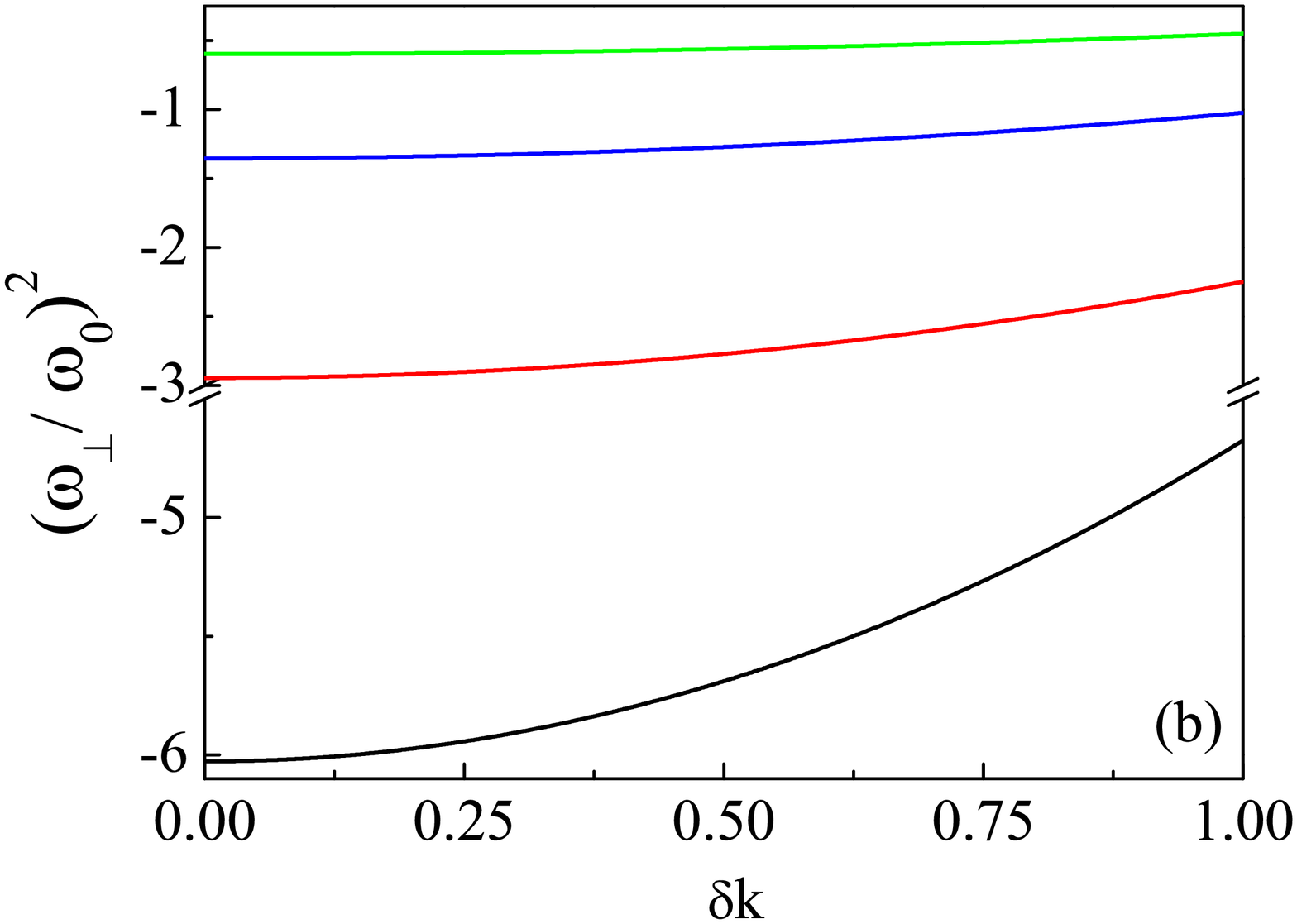}}
\caption{\label{fig:dispersion_dk} (Color online) (a) Dispersion relation $\omega_{\perp}$ as a
 function of $\delta k$ for parabolic confinement (i.e. $\alpha=2$) where dashed, solid and
 dotted lines represent the results for $\upsilon=0.95\upsilon_c$, $\upsilon=\upsilon_c$ and
 $\upsilon=1.05\upsilon_c$ respectively. (b) Square transverse frequency $\omega_{\perp}^2$ as a
 function of $\delta k$ for $\alpha>2$. We have considered $n=1$, $\eta=1$ and different values
 of $\kappa$ as shown in (a).}
\end{center}
\end{figure*}

In Fig. \ref{fig:dispersion_dk}(a) we show the dispersion relation for the normal modes in the
 case of parabolic confinement for different values of $\kappa$ in the three cases, linear regime
 (dotted lines) where the system is stable for any value of the wavevector close to $k_0$, the
 zig-zag regime (dashed lines), and in the transition point (solid lines) where the dispersion is
 linear close to $k_0$.

\subsection{Case $\alpha>2$}
\begin{figure*}
\begin{center}
\includegraphics[width=\textwidth]{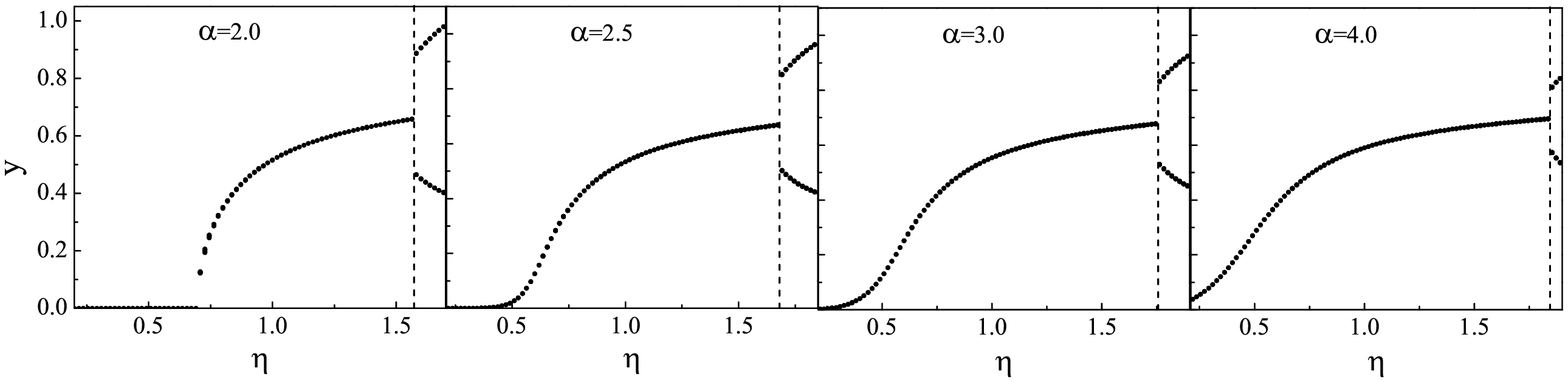}
\caption{\label{fig:mc_alphagt2} Distance of the particles from the $y$-axis as a function of the
 density, for different values of $\alpha$ and $\upsilon=1$. The inter-particle interaction is a
 dipole potential.  The symbols represent the results from Monte-Carlo simulations and the dashed
 vertical lines indicate the transition point from the 2-chains to the 4-chains configuration.}
\end{center}
\end{figure*}

In this case the most simple configuration of the particles is restricted to a 2-chains
 structure, however from Monte-Carlo simulations we know that there is a transition to a 4-chains
 structure after some value of the linear density.  This is shown in Fig. \ref{fig:mc_alphagt2},
 where we plot the distance from the $y=0$ axis of the particles as a function of the density
 considering a dipole inter-particle interaction for different values of $\alpha$.  In those
 figures the 2-chains to 4-chains transition point is marked with a vertical dashed line.  In our
 theoretical model we have found from Eqs. (\ref{omega_perp_general}) and
 (\ref{omega_perp_definition}) that $\omega_{\perp}^2<0$ as shown in Fig.
 \ref{fig:dispersion_dk}(b) and thus the transverse frequency is imaginary and therefore the
 one-chain structure is unstable for any value of the density and the confinement strength.

\begin{figure}
\begin{center}
\includegraphics[scale=0.32]{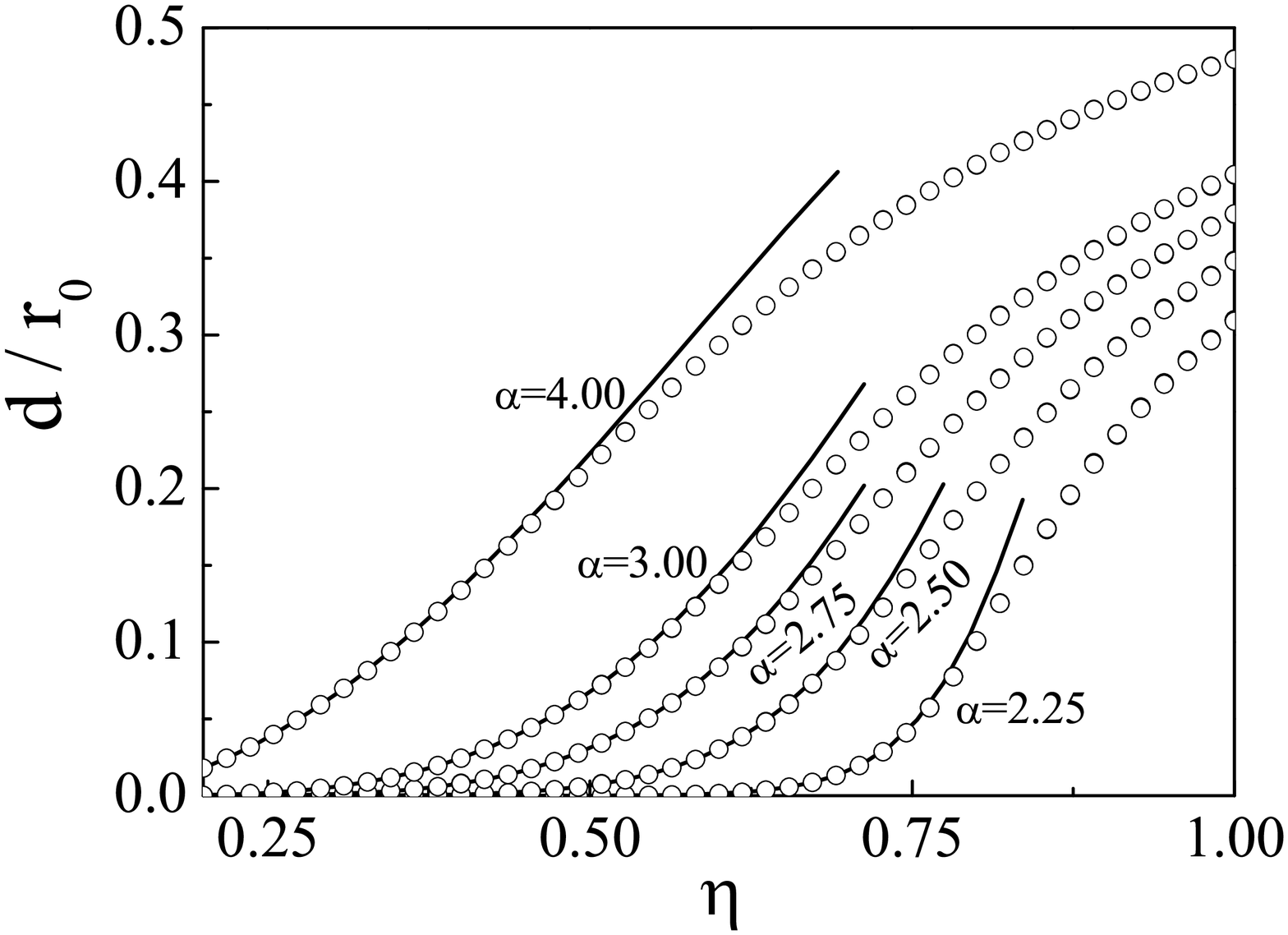}
\caption{\label{fig:c_eta_alphagt2} Order parameter as a function of the linear density in the 
 2-chains region for different values of the exponent of the confinement potential $\alpha$.  The
 solid curves represent the solutions of Eq. (\ref{Ec_euler_lagrange}) and open circles the
 results of our Monte-Carlo simulation. We took the parameters $\upsilon=1$, $\kappa=1$ and $n=1$.}
\end{center}
\end{figure}

This is illustrated in more detail in Fig. \ref{fig:c_eta_alphagt2} where we plot the distance
 of the particles from the $x$-axis for different values of $\eta$.  Note that our mean-field
 results from Eq. (\ref{Ec_euler_lagrange}) agree with the simulation for small values
 of $\eta$.  Note that for small values the confinement potential energy is significantly larger
 than the inter-particle potential energy and therefore the fluctuations of the order parameter
 are smaller.  With increasing $\eta$ the interaction between the particles start to dominant and
 all curves converge to each other (without crossing) for $\eta>1$.

\section{Conclusions}\label{conclusions}

In this work, we studied the critical behavior of a system of particles confined in a 2D
 channel through a $y^{\alpha}$ potential with different functional forms for the inter-particle
 interaction potential.  We derived a Ginzburg-Landau equation for the system and determine the
 behavior of the system close to the transition point where the single chain configuration
 becomes unstable.  We determined the order parameter and its dependence on the external
 confinement and the particle density.

For $\alpha=2$ the critical frequency for the zig-zag transition is larger than for smaller
 values of $\alpha$, which shows that the stability of the linear chain configuration is lower
 for parabolic confinement.  However for low densities ($\eta<1$) the one-chain configuration is
 the most stable state for $\alpha \leq 2$.

For $\alpha>2$ the single chain configuration is unstable for any value of the particle density
 and the strength of the confinement potential.  We found the distance between the two chains as
 function of the particle density. With increasing density a first-order phase transition is
 found to the 4-chains configuration.

For $\alpha<2$ we found analytically no continuous zig-zag configuration irrespective of the
 inter-particle potential.  The instability of the single chain configuration occurs through the
 expulsion of single particles from the chain to $y \neq 0$ positions.

The instability point for $\alpha=2$ is given by $\upsilon_c=\eta_c^{2/(n+2)}$ which becomes an
 almost linear relation, i.e. $\upsilon_c \sim \eta_c$ for $\alpha=1$ and $n=1$.

In a future work we plan to generalize the present analysis to the quantum regime.  for the
 special case of electrons confined by a parabolic potential, i.e. $\alpha=2$, $n=1$ and
 $\lambda=\infty$, such an analysis was presented by J. S. Meyer et. al.
 \cite{038_Meyer, 039_Meng}.  Subsequently the strongly correlated regime which results in
 Wigner crystal physics in quantum wires, was addressed in Ref. [\onlinecite{040_Meyer}].  Such a
 quantum  analysis will address the effect of quantum statistics of the particles and the effect
 of quantum fluctuations on the zig-zag transition.

\section{Acknowledgments}
This work was supported by the Flemish Science Foundation (FWO-Vl).

%

\end{document}